\documentclass[runningheads]{cl2emult}

\usepackage{makeidx}  
\usepackage{graphicx} 
\usepackage{subeqnar} 
\usepackage{multicol} 
\usepackage{cropmark} 
\usepackage{eso}      
\makeindex            

\def\lta{\mathrel{\hbox{\rlap{\hbox{\lower4pt\hbox{$\sim$}}}\hbox{$<$}}}}

\begin{document}

\title*{Structure of the globular cluster M15 and constraints on a
massive central black hole\footnote{To appear in `Black Holes in
Binaries and Galactic Nuclei', Proceedings of an ESO Workshop in
September 1999, Kaper L., van den Heuvel E. P. J., Woudt P.~A., eds.,
Springer-Verlag, 2000.}}

\toctitle{Structure of the globular cluster M15 and constraints on 
a massive central black hole}

\titlerunning{Structure of M15}

\author{Roeland P.~van der Marel}

\authorrunning{Roeland P.~van der Marel}

\institute{Space Telescope Science Institute, 3700 San Martin Drive,\\
Baltimore, MD 21215, USA}

\maketitle


\begin{abstract}
Globular clusters could harbor massive central black holes (BHs), just
as galaxies do. So far, no unambiguous detection of a massive BH has
been reported for any globular cluster. However, the dense
core-collapsed cluster M15 seems to be a good candidate. I review the
available photometric and kinematic data for this cluster. Both are
consistent with a BH of mass $M_{\bullet} \approx 2000 M_{\odot}$,
although such a BH is not unambiguously required by the data. I
discuss some ongoing studies with Keck and HST which should shed more
light on this issue in the coming years.
\end{abstract}

\section{Massive Central Black Holes in Globular Clusters} 

Massive BHs have been convincingly detected in the centers of some
nearby galaxies (e.g., Miyoshi et al.~1995), including our own (Genzel
et al.~1997). In certain galaxies the BH directly reveals itself
through its associated accretion and activity. Such activity is never
observed in globular clusters, but nonetheless, it may well be that
(some) globular clusters also contain BHs. There are many ways in
which globular cluster evolution at high densities (Meylan \& Heggie
1997) can lead to the formation of a massive BH in the center (Rees
1984). For example, core collapse induced by two-body relaxation may
lead to sufficiently high densities for individual stars or
stellar-mass black holes to interact or collide, with a single massive
BH as the likely end product (Quinlan \& Shapiro 1987; Lee 1993,
1995).

The black hole mass $M_{\bullet}$ in galaxies correlates with galaxy
(bulge) mass $M$ such that $M_{\bullet} / M \approx 10^{-3 \pm 1}$
(Kormendy \& Richstone 1995; Magorrian et al.~1998; van der Marel
1999). One may use this correlation to obtain a crude estimate for the
possible BH masses in globular clusters (although it should be kept in
mind that the BHs in galaxies are often hypothesized to have formed
through gas collapse and accretion, instead of through collapse of a
stellar cluster; e.g., Haehnelt et al.~1997). This yields $M_{\bullet}
\approx 10^3 M_{\odot}$.

The presence or growth of a BH in a globular cluster affects both the
stellar density profile and the stellar dynamics. Observational
constraints on BHs in globular clusters can therefore be found from
photometric and kinematic studies. So far, no detection of a massive
BH has been obtained for any globular cluster, but only few, if any,
studies have had sufficient sensitivity to unambiguously detect BHs
with masses as low as $M_{\bullet} \approx 10^3 M_{\odot}$. On the
other hand, advances in observational capabilities and techniques are
now making it possible to study BH masses down to this limit, so this
is becoming a more active area of interest.

The globular cluster M15 (NGC 7078) at a distance of 10 kpc is one of
the densest globular clusters in our Galaxy. The presence of a bright
X-ray source (Hut et al.~1992) and several millisecond pulsars
(Phinney 1993) are manifestations of its extreme density, which makes
M15 one of the best a~priori candidates to search for evidence of a
central BH. For this reason, M15 has been intensively observed in the
past decade using a variety of techniques, and it is also the subject
of several ongoing studies.

\section{M15 Photometry}

Core-collapsed clusters have stellar surface density profiles that
rise all the way into the center. Such clusters make up $\sim\! 20$\% of
all globular clusters in our Galaxy, and stand in marked contrast to
King-model clusters, which show flat central cores and are modeled as
tidally-truncated isothermal systems. At ground-based resolution M15
has long been known as the proto-typical core-collapsed cluster
(Djorgovski \& King 1986; Lugger et al.~1987). Studies with the Hubble
Space Telescope (HST) in the past decade have provided significantly
higher spatial resolution than ground-based data, but have also not
provided any evidence for a homogeneous core in M15 (despite early
claims to the contrary; Lauer et al.~1991).

Guhathakurta et al.~(1996) used the HST/WFPC2 to image M15. Individual
stars were resolved well below the main sequence turnoff even in the
dense central few arcsec. The projected surface number density profile
between $0.3''$ ($0.017$ pc) and $6''$ (after correction for the
effects of incompleteness and photometric bias/scatter) was found to
be well represented by a power law $N(R) \propto R^{-0.82 \pm
0.12}$. While the density profile cannot be measured reliably at radii
$\lta 0.3''$, mainly because of small-number statistics and
uncertainties in the position of the cluster center, there is no
evidence from non-parametric studies that the density profile would
flatten at smaller radii.
 
Sosin \& King (1997) obtained even higher spatial resolution data with
the HST/FOC. They find $N(R) \propto R^{-0.70 \pm 0.05}$ for turnoff
stars, consistent with the Guhathakurta et al.~analysis, and show in
addition that the power-law slopes are slightly different for stars of
different masses. This is qualitatively consistent with the mass
segregation predicted in a cluster in which two-body relaxation has
been important.

Bahcall \& Wolf (1976, 1977) constructed detailed models for the
equilibrium stellar density distribution of a globular cluster in
which a BH has been present for much longer than the two-body
relaxation time. For a cluster of equal-mass stars one expects $N(R)
\propto R^{-3/4}$, in surprisingly good agreement with the star count
profile for M15. However, the observed profile can be explained
equally well as a result of core-collapse (Grabhorn et al.~1992; but
note that the predictions from Fokker-Planck models are quite
uncertain). Hence, the star count profile by itself yields only
limited insight. An additional problem is that photometric studies
cannot determine whether light follows mass, and what the abundance
and distribution of dark remnants are. Kinematical studies are
therefore essential to gain further insight.

\section{M15 Kinematics}

M15 has been the subject of many ground-based kinematical studies.
Observational strategies have focused primarily on the determination
of velocities of individual stars, using either spectroscopy with
single apertures, long-slits or fibers (Peterson, Seitzer \& Cudworth
1989; Dubath \& Meylan 1994; Dull et al.~1997; Drukier et al.~1998),
or using imaging Fabry-Perot spectrophotometry (Gebhardt et al.~1994,
1997, 2000). Integrated light measurements using single apertures have
also been attempted (Peterson et al.~1989; Dubath, Meylan \& Mayor
1994), but are only of limited use; integrated light spectra are
dominated by the light from only a few bright giants, and as a result,
inferred velocity dispersions are dominated by shot noise (Zaggia et
al.~1992; Dubath et al.~1994).

Line-of-sight velocities are now known for $\sim\! 1800$ M15 stars, as
conveniently compiled by Gebhardt et al.~(2000). The projected
velocity dispersion profile inferred from this sample is shown in
Fig.~1. It increases monotonically inwards from $\sigma = 3 \pm 1$
km/s at $R=7$ arcmin, to $\sigma = 11 \pm 1$ km/s at $R=24''$. The
velocity dispersion is approximately constant at smaller radii, and is
$\sigma = 11.7 \pm 2.8$ km/s at the innermost available radius $R
\approx 1''$ (Gebhardt et al.~2000). The figure also shows the
predictions of spherical dynamical models for M15 in which the
velocity distribution is isotropic and the stellar population has a
mass-to-light ratio $M/L = 1.7$ (in the V-band) that is independent of
radius. Different curves correspond to different BH masses. A BH
causes the velocity dispersion to rise in Keplerian fashion as $\sigma
\propto R^{-1/2}$ towards the center of the cluster. The best-fitting
model of this type has $M_{\bullet} \approx 2000 M_{\odot}$. However,
it should be noted that the data can be fit equally well with a model
in which the $M/L$ of the stellar population increases inwards to a
value of $M/L \approx 3$ in the center. This would not be implausible,
since mass segregation would tend to concentrate heavy dark remnants
to the center of the cluster.

\begin{figure}
\centering
\includegraphics[width=.5\textwidth,angle=270]{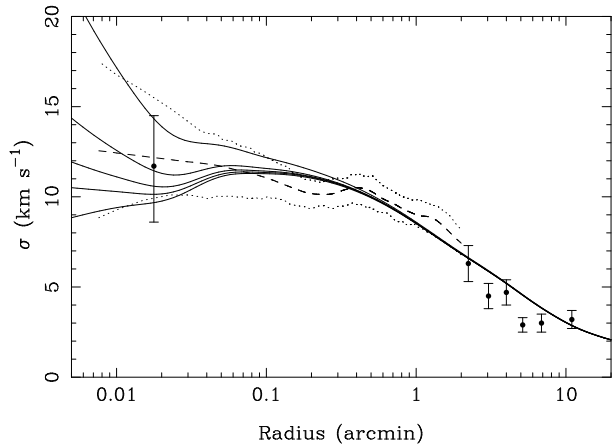}
\caption[]{Projected velocity dispersion profile of M15. Binned data
are shown as dots with error bars, while a non-parametric estimate of
the dispersion profile in the central arcmin is shown as a dashed line
bounded by two dotted lines (the latter representing the 68.3\%
confidence region). Solid curves are predictions of isotropic models
with BHs that have $M_{\bullet} = 0, 500, 1000, 2000,$ and $6000
M_{\odot}$, respectively. (Figure reproduced from Gebhardt et
al.~2000).}
\label{fig1}
\end{figure}
 
Phinney (1993) used an interesting alternative argument to constrain
the mass distribution of M15. There are two millisecond pulsars in M15
at a distance $R = 1.1''$ from the cluster center that have a negative
period derivative ${\dot P}$.  This must be due to acceleration by the
mean gravitational field of the cluster, since the pulsars are
expected to be spinning down intrinsically (positive ${\dot P}$). The
observed ${\dot P}$ values place a strict lower limit on the mass
enclosed within a projected radius $R = 1.1''$. Combined with the
observed light profile this yields $M/L > 2.1$ for the total
mass-to-light ratio within $R \leq 1.1''$, with a statistically most
likely value of $M/L \approx 3.0$. These results are consistent with
the stellar kinematical analysis, and also indicate that there must be
some increase in $M/L$ towards the cluster center (since at large
radii $M/L \approx 1.7$). However, also the pulsar data cannot
discriminate whether this is due to mass segregation or a central BH.

It has been known for some time that M15 has a net global rotation
amplitude of $V \approx 2$ km/s. This is somewhat surprising given the
short relaxation time, which tends to isotropize the velocity
distribution. However, more recent work (Peterson 1993; Gebhardt et
al.~2000) has revealed two even more puzzling facts. First, the
position angle of the projected rotation axis in the central region is
$\sim\! 100^{\circ}$ different from that at larger radii. Second, the
rotation amplitude increases to $V = 10.4 \pm 2.7$ km/s for $R \lta
3.4''$, so that $V / \sigma \approx 1$ in this region. Although the
central increase in rotation amplitude may have something to do with
the presence of a BH, neither of these observations fits naturally in
any current theory of globular cluster structure.

\section{Ongoing Studies and Future Prospects}

To make further progress in our understanding of the structure of M15,
and in particular to determine whether it harbors a central BH or not,
it is of crucial importance to obtain more stellar velocities close to
the center. However, velocity determinations at $R \lta 2''$ are very
difficult due to severe crowding and the presence of a few bright
giants in the central arcsec. Gebhardt et al.~(2000) used an Imaging
Fabry-Perot spectrophotometer with adaptive optics on the CFHT, and
obtained FWHM values as small as $0.09''$. However, the Strehl ratio
was only $\lta 6$\%, so that even in these observations the light from
the fainter turnoff and main-sequence stars in the central arcsec was
overwhelmed by the PSF wings of the nearby giants. As a result, there
are only 5 stars in the central arcsec with known velocities.

In an attempt to improve this situation my collaborators and I
initiated two new observational studies. In the first (Guhathakurta et
al., in progress) we used the HIRES echelle spectrograph on the Keck~I
telescope in multislit mode. The seeing FWHM was $0.7''$ and the slit
width $0.5''$. These observations will not allow us to spatially
resolve stars in the central arcsec spatially, but the high spectral
resolution (as compared to Fabry-Perot spectrophotometry) may allow us
to resolve stars spectrally. In the second, more ambitious study we
are using the STIS long-slit spectrograph on HST to map the center of
M15 spectroscopically (Cycle 8, van der Marel et al., in progress). We
are stepping a $0.1''$-wide slit across the center in 23 steps of
$0.1''$. This will yield a spectrum of each $0.1'' \times 0.1''$ cell
in a rectangular grid on the center of M15. The spectra are taken
around the Mg b triplet. The expected signal-to-noise ratio should be
sufficient to extract stellar velocities using cross-correlation
techniques for several tens of faint stars in the central few
arcsec. This will improve our knowledge of the velocity dispersion
profile and will put new constraints on the possible presence of a
BH. Also, the central escape velocity of M15 in the absence of a BH is
expected to be $\sim\! 40$ km/s (e.g., Webbink 1985); any (non-binary)
stars found to have velocities exceeding this value will provide
additional and independent evidence for a central mass concentration.

More progress in the near future may come from the availability of
stellar proper motion measurements with HST. Several groups are
pursuing this, both for M15 and for other clusters. The positional
accuracy that can be achieved with the HST/WFPC2 is $\sim\! 0.01$ PC
pixel ($0.5$ mas). For a 5 year baseline, motion over this distance
corresponds to $5$ km/s (at the distance of M15). This opens up the
exciting prospect of having all three velocity coordinates for a large
sample of stars. Whether proper motions can be determined all the way
into the central arcsec remains to be seen though, since the severe
crowding will complicate positional measurements in that region.

\section{Conclusions and Acknowledgements}

Globular clusters could have central BHs, and M15 is the best
candidate so far. The available photometry and kinematics are
consistent with a BH of mass $M_{\bullet} \approx 2000 M_{\odot}$,
although such a BH is not unambiguously required by the data. Ongoing
studies should shed more light on the structure of M15 in the coming
years.

I am grateful to my collaborators, Raja Guhathakurta, Ruth Peterson,
Pierre Dubath, Karl Gebhardt and Tad Pryor for many stimulating
discussions on this subject.


\end{document}